# A HYBRID MACHINE LEARNING MODEL TO STUDY UV-Vis SPECTRA OF GOLD NANO SPHERES

B. KARLIK[1], M. F. YILMAZ[2*] M. OZDEMIR[3], C.T. YAVUZ[4], Y. DANISMAN[5]

[1]Neurosurgical Simulation and Artificial Intelligence Learning Centre, Montreal Neurological Institute, McGill University, H3A2B4, Montreal, QC, CANADA

[2]Independent Researcher, Fremont, CA, 94538, USA

[3]Basic Sciences, College of Engineering, Imam Abdulrahman Bin Faisal university, P.O. Box 1982, Dammam, 31451, KSA

[4]Korea Advanced Institute of Science and Technology, 291 Daehak-ro, Yuseong-gu, Daejeon 34141, Republic of Korea

[5]Department of Mathematics and Computer Sciences, Queensborough Community College, CUNY, Bayside, NY, USA

*Correspondence to fthyilmaz53@gmail.com

## ABSTRACT

Here, we have employed Principal Component Analysis (PCA) and Linear Discriminant Analysis (LDA) to analyze Mie calculated UV-Vis spectra of gold nanospheres (GNS). Eigen spectra of PCA perform the Fano type resonances.3D vector field spectra reveal the Homoclinic orbit Lorenz attractor. Quantum confinement effects are observed by 3D representation of LDA. Standing wave patterns resulting from oscillations of ion acoustic phonon and electron waves are illustrated through the eigen spectra of LDA. Such capabilities of GNPs have brought high attention for the high energy density physics applications. Furthermore, accurate prediction of gold nanoparticle (GNP) sizes using machine learning could provide rapid analysis without the need for expensive analysis. Two hybrid algorithms consist of unsupervised PCA and two different supervised ANN have been used to estimate the diameters of GNPs. PCA based artificial neural network (ANN) were found to estimate the diameters with a high accuracy.

## INTRODUCTION

The arrangement of metal nanoparticles in an order of chain or array provides surface plasmon polaritons (SPPs), which are the bosonic (quasiparticles) coupled modes of an incident electromagnetic wave and free electrons (plasmon) of surface. SPPs can manipulate the electromagnetic radiation and perform as micro-optical devices and can propagate in undistorted manner for several diffraction lengths along a metal surface [1,2 and 3]. Fano resonance stands as the primary physical mechanism behind the manipulation of electromagnetic waves. Fano resonances are generated by the coupling of discrete excited states to a continuum of states and can be produced through the scattering of the electromagnetic waves by nano-particles [4 and 5 ]. Furthermore, Semouchkina et al., showed that Fano-type shape, and significant strength of the observed resonance are characteristic of the interaction of dielectric resonator arrays with Fabry-Perot standing waves [6]. Besides, Sergeyev et al., 2014 proposed that spiral attractors generated by the interaction of dissipative solitons with carbon nanotubes are the main mechanisms which provide the trapping and manipulation of light [7] Such capabilities of NPs brought attention of application for high energy density physics. Kaymak et al., showed that ultra- dense high energy density plasmas ($n_e > 9x10^{24}$ $cm^{-3}$) can be generated by the irradiation of relativistic intensity femtosecond laser with aligned nanowires [8]. Ostrikov et al., proposed that several important properties of nano structured targets are important for various applications such as creation of dense hot plasmas with a temperature of 2-4 keV, efficient x-ray and ion sources. [9]. Lastly, Rocca et al., 2018 has justified that array formatted nanowires completely absorb the high energy pulses and thereby producing efficient number of neutrons [10]. These capabilities of nanoparticles have strong impacts in the field of high energy density physics of fusion applications.

UV-Vis spectroscopy as a useful tool is employed for the characterization, estimation of the sizes of nanoparticles, concentration and aggregation level. It is a useful technique as UV-vis spectrometers can be found in many laboratories, the analysis does not change the sample, and the time needed for registration of the spectrum is very short. If the appropriate correction of the metal dielectric constant for the nanoparticle size and physicochemical environment is provided, then Mie theory can be used to analyze the extinction spectra of AuNP recorded by UV-vis spectroscopy [10,11 and 12]. On the other hand, Yilmaz et al., 2019 showed that application of PCA and LDA over the spectral database of plasma could inform more physical insight of deeper structures and polarization types of the ion species and electron ion oscillations in the plasma [13 and 14].

PCA and LDA are the two of the most common pattern recognition techniques that are used in order to reduce the dimensions of a given raw data in pattern recognition problems. PCA uses rotational transformation in a way that most of the data variability remains in a space of low dimensions, and it ignores the remaining dimensions that contain little variability. LDA, as a pattern recognition technique, is quite similar to that of PCA. The main difference is that LDA determines the vectors that best separate the classes while trying to keep the variance maximum. Although it looks like LDA outperforms PCA in multi-class settings where class labels are known, it might not always be the case, especially if the sizes of the classes in the data sets are relatively small. Besides, LDA performs well only if the classes have equal co-variance. PCA is good at keeping dimensions of highest variance, but it can disregard discriminant dimensions where LDA is needed. It is easy to find examples where LDA outperforms PCA and vice versa [13 and 14].

Artificial neural network (ANN) algorithms have also been of great interest and used over the last couple of decades in many applications and areas such as remote sensing, computer vision, pattern recognition and medical diagnosis. The ANN algorithms can easily identify and extract the patterns by setting correlation between sets of given inputs and outputs through a training process. This adaptive nature makes them particularly appropriate for solving complex and non-linear problems. Another advantage of a machine learning algorithm is the fact that it can establish strong correlation between the parameters without any knowledge of them. Therefore, it enables the handling of uncertainties, data with noise and nonlinear relationships which are hard to determine [15].

ANNs have been applied to both to X-ray spectra to predict plasma electron temperatures and densities, and to UV- Vis and FTIR spectra to investigate forensics [16 and 17]. Lately, they have been used to estimate the diameters of nanoparticles [18 and 19]. Peurifoy et al., have proposed artificial neural networks to approximate light scattering by multilayer nanoparticles [20].

Selection of inputs is one of the most significant components of designing classifier based upon pattern recognition since even the best classifier will perform badly if the inputs are not chosen very well. The complexity of classifier method is inversely proportional to its classification efficiency. Recently, Karlik et al., has indicated the positive effects of hybrid learning methods consisting on unsupervised clustering and supervised classifiers [21]. So, in this work, a hybrid classifier consisting of an unsupervised PCA and a supervised ANN algorithm has used. In this hybrid classifier, the input nodes of supervised ANN activations are derived through unsupervised PCA of the input data, so that the neural system could deal with the statistics of the measurement error directly. Therefore, we have first employed Mie scattering generated spectra of GNS as our database. This database was then analyzed by PCA and LDA methods, and the physical interpretations of PCA and LDA of UV-Vis spectra of gold were examined. Then, the extracted PCA coordinates were used for the training of ANN to estimate the diameters of GNS. Finally, the experimental spectra of gold spheres with diameters 5, 7, 10, 15, 20 and 30 nm were tested for the estimation of diameters. Our findings reveal that PCA based artificial neural network (ANN) estimates the diameters of gold nanoparticles with a remarkably high accuracy.

## METHODS

### Materials

Hydrogen tetrachloroaurate (III) trihydrate ($HAuCl_4.3H_2O$, gold salt), trisodium citrate (reducing agent), and sodium chloride were obtained from Aldrich and used as received. Gold nanoparticles (AuNPs) with the sizes of 5, 7, 10, 15, 20 and 30 nm (Fig.1) were purchased from Nanocomposix, Inc. Distilled water was used to prepare the aqueous solutions of AuNPs. Au nanoparticles were synthesized through the well-established reduction methods of Turkevich, Frens and Jana [22, 23 and 24]. In the preparation of 5 nm AuNPs, a 10 mL aqueous solution containing 0.25 mM $HAuCl_4$ and 0.25 mM trisodium citrate were first prepared. Ice-cooled 0.3 mL of 0.1 M sodium borohydride was added to this solution while stirring. The color change in the solution was an indication of the particle formation.

### Synthesis

For the synthesis of 7 nm or larger AuNPs, aqueous solutions of 0.254 mM $HAuCl_4$ and 38.8 mM trisodium citrate were first prepared. In a typical synthesis, 50 mL $HAuCl_4$ solution was heated to boiling. After 5 min., 0.4-2.0 mL trisodium citrate solution was added to this mixture at once and the mixture was stirred for approximately 15 min. The color of the reaction mixture turned from yellow to colorless and then ruby red color depending on the AuNPs sizes. After cooling to RT, the samples were centrifuged and washed several times with DI water to obtain 5, 7, 10, 15, and 30 nm citrate-coated AuNPs.

### Characterization

Dynamic light scattering (DLS, Malvern Zetasizer Nano ZS) technique was used to determine the average hydrodynamic diameters of gold nanoparticles. Nano ZS detects the scattered light at an angle of 173°, known as backscatter detection, by using He–Ne laser (4 mW) operated at 633 nm. Aqueous 10 mM NaCl solutions were used for all DLS measurements.

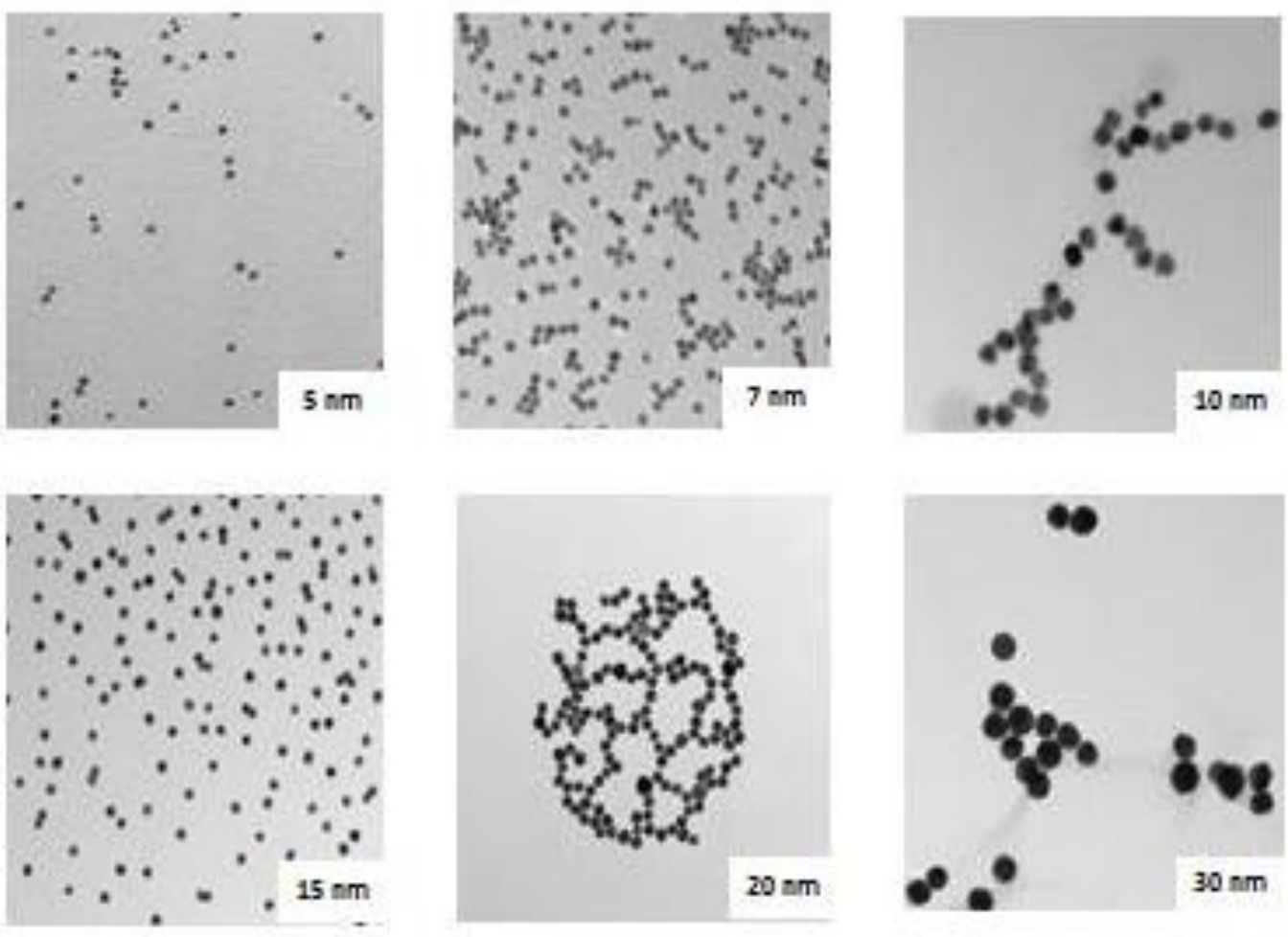

**Fig. 1.** TEM images of gold nanoparticles

**Mie Scattering**

The Mie theory describes the scattering of an electromagnetic wave by homogenous spherical particles. Although it is based on idealized initial conditions, it is widely used for the radiation problems in a light scattering media. This theory mainly calculates the coefficients for absorption, scattering and extinction. One can find several programs that can perform Mie theories-based solutions. The main advantage of the Mie methods is that they suggest solutions for the cases where the diameter of scattering particle is comparable to the wavelength of the light. For much larger or smaller particle sizes, there are already several simple methods that can be used in order to describe the behavior of the corresponding systems. Since our focus is on the sizes similar to the incoming light wavelength, we employed Mie theory calculated spectra to create our database [12 and 25]. Our database contains 25 spectra of gold spheres without polymer shell with the diameters 2, 4, 6, …, 50 nm. In Fig.2, experimental spectra of 5, 7, 10, 15 and 30 nm and their Mie calculated spectra are illustrated. By applying PCA to the whole database, eigenvalues with corresponding eigenvectors of the covariance matrix are obtained. Throughout this work, only the first three eigenvectors, which correspond to the largest three eigenvalues, are considered to reduce the dimension of the initial.

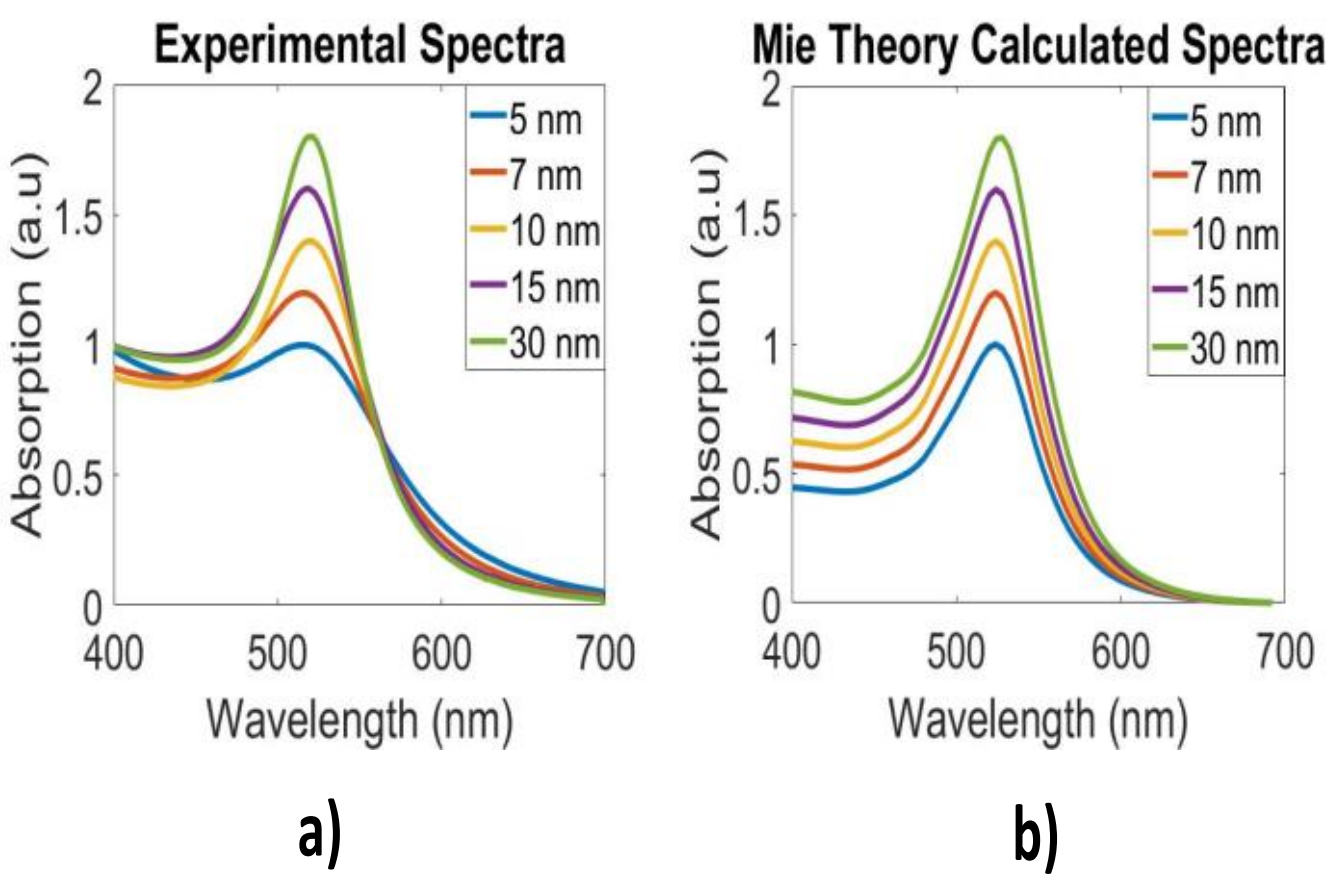

**Fig. 2.** a) Experimental UV-Vis spectra and b) Mie calculated spectra

The Mie theory we used in this work employs the spherical vector harmonics to express the electromagnetic fields in the form of scattered shape for the dipole interactions. Vector-basis functions, are derived from the separable solutions to the scalar Helmholtz equation. The relative intensity of the scattering or absorption of particles with different sizes is described by the particle efficiency which *is* calculated by dividing the scattering/absorption/extinction cross section by the geometrical cross-sectional area. These cross sections are obtained by the integration of the Poynting vector for a spherical radius. The experimental size dependent dielectric electric effects in gold nanoparticles is described by a Drude conduction model. In addition, enhancement of the local intense fields due to the presence of local surface roughness are described by amplification mechanisms in the model [12 and 25].

**Principal Component Analysis (PCA)**

The main purpose of PCA is to find a subspace spanned by the vectors with largest variances. This is an optimization problem, which reduces the dimension of a data set while retaining the variance. It is performer by converting the data set of possibly correlated variables into linearly uncorrelated variables called Principal Components (PC). These PCs determine the similarities and differences of the data, and they are the eigenvectors of the covariance matrix which consists of covariances of all different dimensions. Hence, every element of the original data can be written as a linear combination of PCs. To reduce the dimension, only the eigenvectors (|PC1>, |PC2> and |PC3> in this paper) which correspond to the largest (dominant) eigenvalues of the covariance matrix are considered. The original data is projected into the space spanned by these PCs which are orthonormal. In this way, some information is lost due to not considering the eigenvectors corresponding to the small eigenvalues, but this information has less significance [13].

**Linear Discriminant Analysis (LDA)**

Similarly, to PCA, LDA is another dimension reduction technique to identify the hidden structures of large data sets [14]. LDA is applied to data sets, which consist of different classes of similar elements and used to find vectors, to discriminate the classes while respecting the similarities among the class members. Unlike LDA, it deals with the entire data, and it does not consider different classes. Therefore, LDA is applied to data sets when different classes must be considered. The eigenvectors corresponding to the largest three eigenvalues in LDA are |LD1>, |LD2> and |LD3>.

**Artificial Neural Networks**

Artificial Neural Network (ANN) is the most popular and useful method of machine learning algorithms. One of main features of ANN is the possibility to adapt their behavior to the changing characteristics of the modeled system. As a parallel processing distributed system, ANN depends on learning through a training set of data using a supervised learning algorithm [26]. The processing units in feed forward and back-propagation neural networks are arranged in multi layered perceptron (MLP) architectures which have back-propagation (BP) algorithm and uses various activation functions. Most commonly used non-linear activation functions are the sigmoid and the hyperbolic tangent functions [27 and 28]. Other difficulties of deep learning for feed-forward neural networks with multilayer perceptron are included decision how many numbers of nodes of hidden layers, optimum selection of momentum coefficient and learning rate [29 and 30]. As seen in Fig. 3, each layer of MLP is fully connected to the previous layer and has no other connections. The MLP consists of 3 or more layers including one input, one output and one or more hidden layers. Multiple hidden layers of non-linearly activating nodes make a deep neural network. In this study, we used four layered MLP architecture; input and output layers with one node, and hidden layers with five nodes.

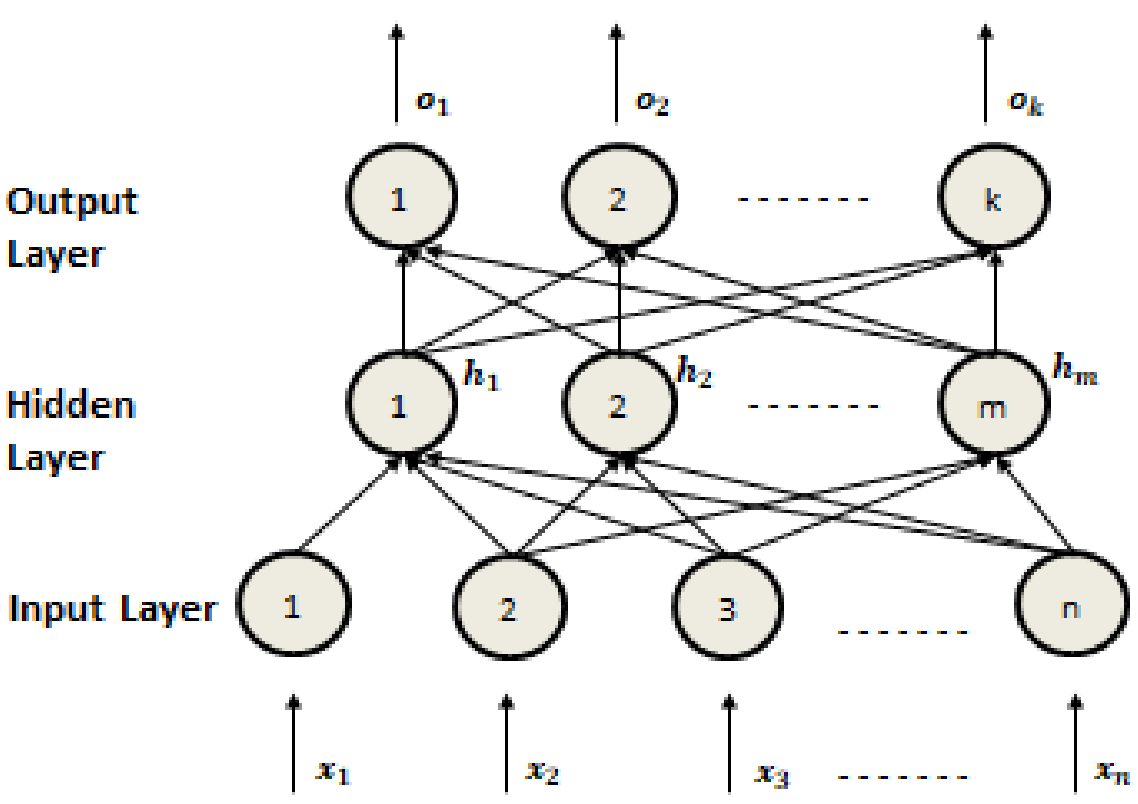

**Fig. 3.** Illustration of a general MLP architecture.

The BP with generalized delta learning rule with an iterative gradient algorithm is implemented to minimize the mean square error (MSE) between the actual output of a multilayered feed-forward neural network and a target output. MSE is also used to measure how well ANN works.

In this study, two different MLP architectures (ANN-1 and ANN2) were used. The architecture of ANN-1 is 1:25:1 which has only 1 hidden layer. The other is 1:5:5:1 which has 2 hidden layers. After trying several numbers of hidden nodes for both MLP architectures, we have selected optimum number as 25 and 5 respectively. Single input is usually a significant problem in MLP architecture. Moreover, the learning rates and momentum coefficients are chosen to be 0.1 and 0.95 respectively after trying many numbers of coefficients.

## RESULTS and DISCUSSION

Fig. 4a illustrates the 3D representation of PCA coordinates of GNS, which exhibits a parabolic profile. Fig. 4b illustrates the mean absorbance and vector representation of the spectra. |PC1> vector represents asymmetric line shape of Fano like resonance. Fano resonance which results from the interference of the scattering amplitudes of continuous (bright mode) and discrete states (dark mode) provides a field enhancement. It is known that non-diffractive (airy) beam of surface Plasmon polariton (SPP) wave follows parabolic profile during its travel and Fano resonance originates from the coupling of SPP and waveguide modes [31,32,33 and 34]. Therefore, Fig. 4 shows that PCA can efficiently extract the propagation and resonance characteristics of the SPP over spectral database of GNS. Furthermore, SPPs are observed in the array and chain form of nanoparticles. Since our database is generated by the spectra of different diameters of gold nanospheres in ascending order, such an array behavior is expected [35].

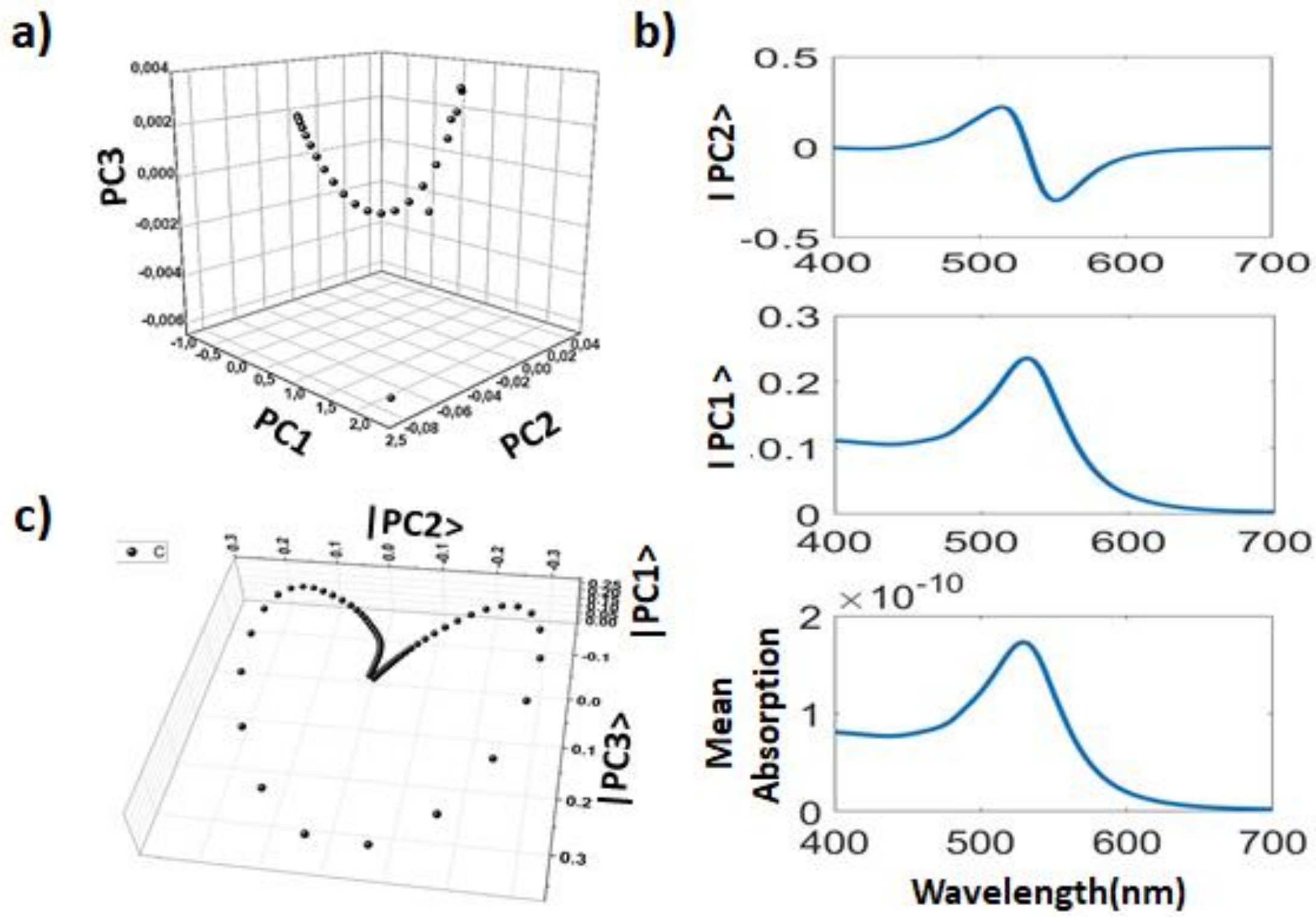

**Fig. 4.** a) PC1, PC2 and PC3 coefficients b) Spectra of the mean of the initial data, and first two dominant eigenvectors IPC1> and IPC2> of the covariance matrix, c) 3D vector field of |PC1>, |PC2> and |PC3>.

3D vector spectra plot is given in Fig.4c which represents the homoclinic orbit with saddle point. The homoclinic orbit is the one of the main examples of the occurrence of chaotic strange attractors observed 3D vector fields [36]. Such attractors arise under bifurcations of resonant points (fixed saddle point). Luk'Yanchuk et al., stated that Fano resonances from light scattering by nanowires is accompanied by bifurcations of the Poynting vector[37]. Whistler mode plasma waves are also an example of monoclinic orbit attractors generated by stimulated Raman and Brillouin scattering. The process is the conversion of an incident photon into a forward and backscattered photon and ion acoustic waves (phonons) [38].

In Fig.5a LDA coordinates in 3D exhibit quantum confinement structure. Such confinements are expected as the size of the particle decreases to nanoscale comparable to the electron's wavelength. The electrons in these structures behave like a particle in potential well. Confined standing waves are the time independent solutions of the Schrodinger equations in the potential well which are formed by concurring of two anti-propagating surface plasmon waves [39]. Fig. 5b shows the LDA vector spectra which illustrate the oscillating wave pattern of ion acoustic waves (|LD1>) and electrons (|LD2>) [14]. Fast Fourier transform modeling of ion acoustic and electron waves results in oscillation frequency of 31 Hz and 72 Hz, respectively. For reference, typical dust phonons (visible ion acoustic) waves oscillate around 5 -35 Hz [40 and 41].

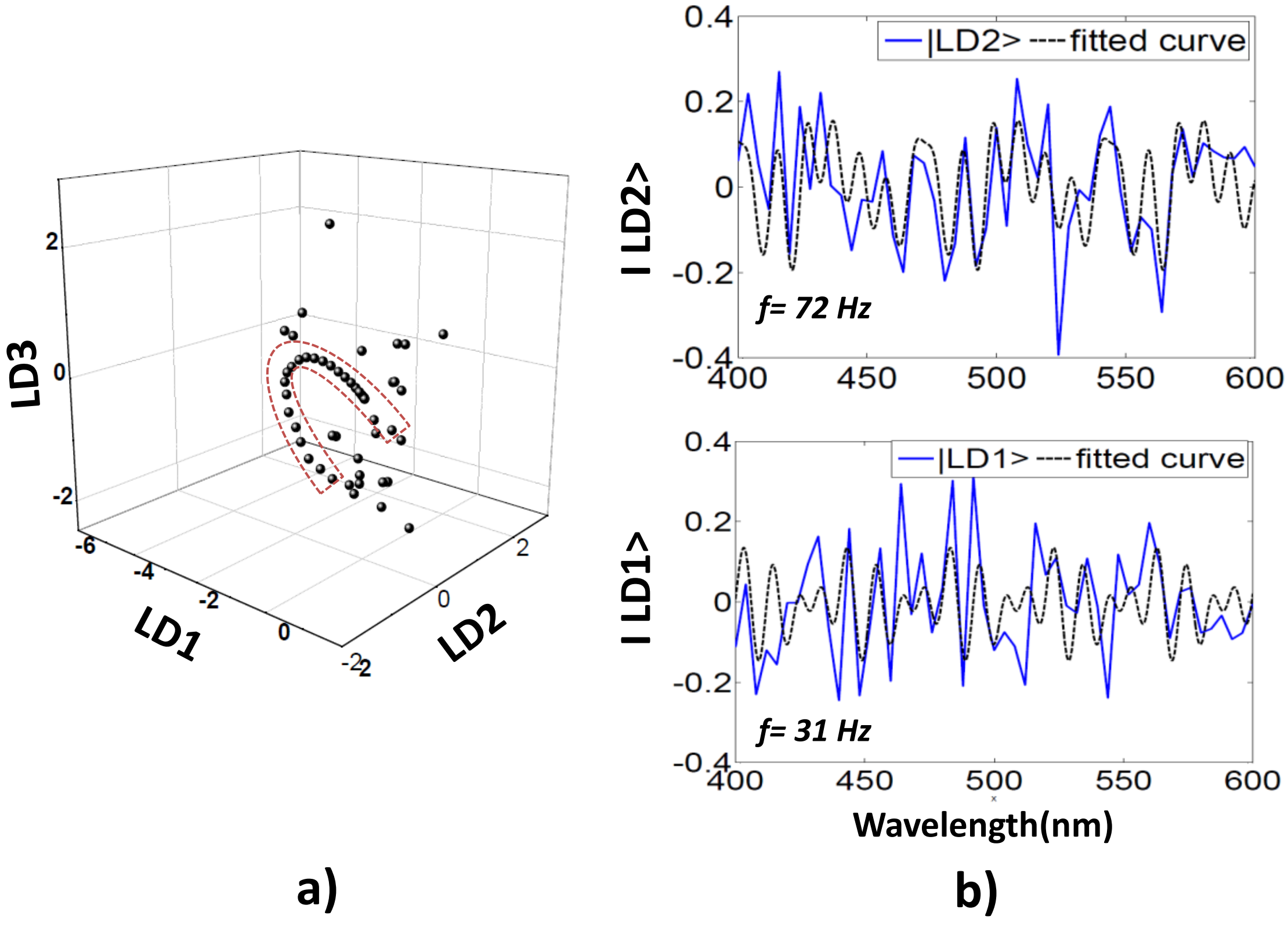

**Fig. 5.** a) LD1, LD2 and LD3 coefficients b) The spectra of the first two dominant eigenvectors: ion acoustic (ILD1>) and electron (ILD2>) waves obtained in LDA.

Fig.6a shows that there is a nonlinear relation between the extracted PCA and LDA coordinates and the corresponding diameters. PCA coordinates which are obtained from the spectra produced from Mie scattering theory are used for the training process, and the outputs are the diameters of the nanoparticles. The coordinates of PCA of Mie calculated spectral database are used for the training of ANN which estimates the diameter of the GNP with high accuracy. In this setting, it should be pointed out that PCA method provides an advantage for the use of ANN and the other machine learning algorithms as it enables to reduce the dimension of inputs. Since LDA of spectra of GNSs reveals structures of nonlinear ion and electron oscillations (Fig. 6b), LDA based ANN was not studied in this work.

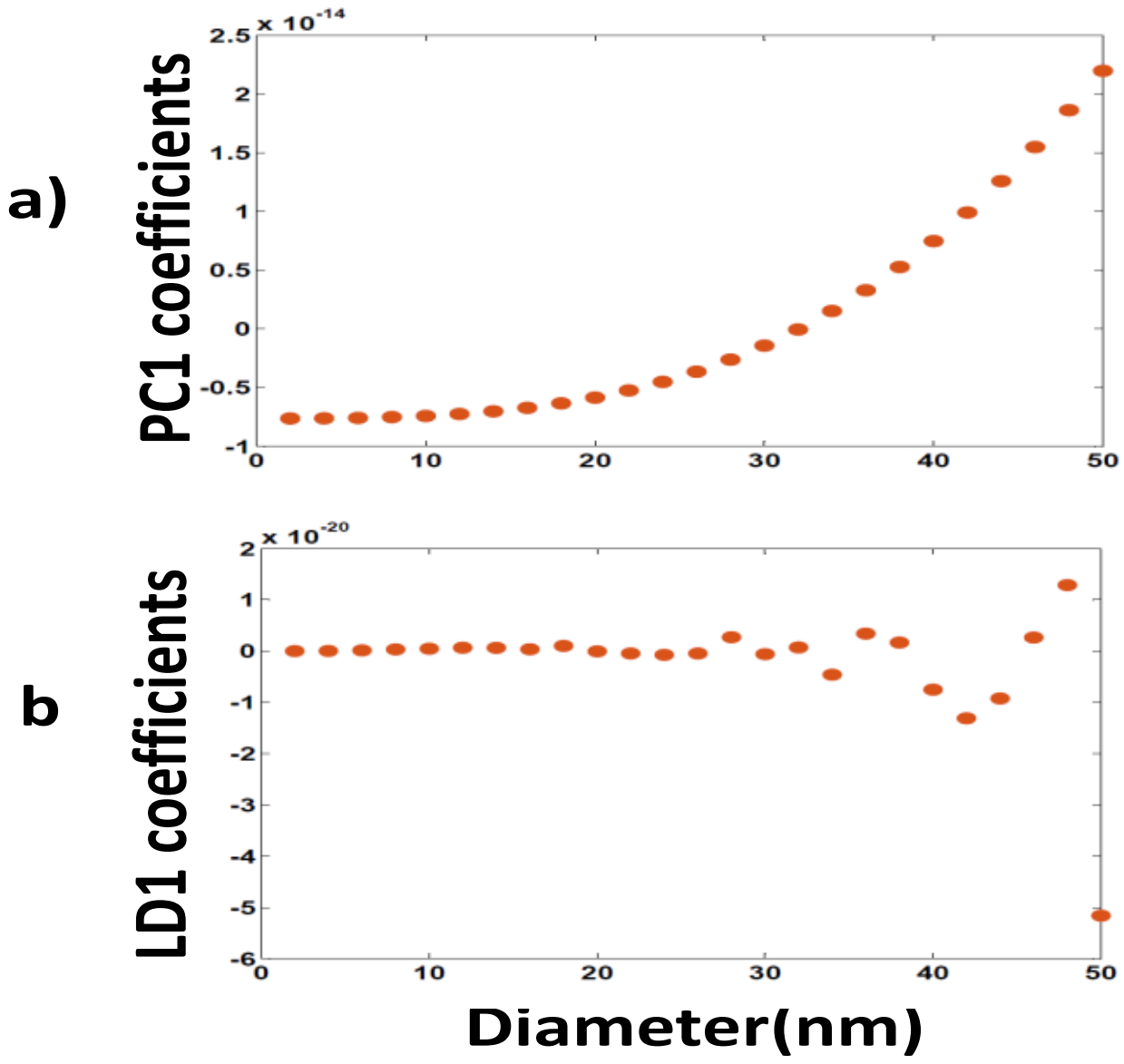

**Fig. 6.** Plot of a) |PC1> and b) |LD1> coordinates versus GNS diameters

In Table 1, Dynamic Light Scattering (DLS) measurements of gold spheres and estimations of diameters obtained by ML algorithms are compared. ANN-2, which consists of (1:5:5:1) MLP structure has shown the best recognition accuracy which is 100%. Additionally, ANN estimations show that average testing zero error was found for 6 different diameters.

| NSP (nm) | DLS Diameter (nm) | ANN-1 (nm) | ANN-2 (nm) |
|---|---|---|---|
| 5 | 5.0±0.6 | 7 | 5 |
| 7 | 7.5±0.8 | 8 | 7 |
| 10 | 11.6±1 | 9 | 10 |
| 15 | 15.4±1.5 | 13 | 15 |
| 20 | 20.5±2.1 | 20 | 20 |
| 30 | 31.0±3 | 30 | 30 |
| MSE_Test_Data | | 1.21 | 0 |

Table 1. Values of measured and estimated diameters with error.

It is difficult to estimate 1I1O (one input-1 output) complicated data even if it has small-scale data which has complex convex region on the data space according to data mining rules. It is also difficult to estimate the spectra for small nanoparticles of different sizes. In this work, we have solved this problem by using multi hidden layered ANN as a deep learning algorithm. Deep learning is a class of machine learning algorithms that uses a cascade of many layers of nonlinear processing units for feature extraction and transformation as in ANN-2. Each successive layer uses the output from the previous layer as an input.

In a similar study, Peurifoy et al. 2018 have found that test results are better than validation results for the nano particular shell 2 and 3, and for the others are the same percentages. These are abnormal results according to ANN because the validation set is different from test set. The validation set can be considered a part of training set because it is used to build ANN mode. However, it is usually used for parameter selection and to avoid overfitting. If the used model is non-linear (like ANN) and trained on a training set only, it will most likely to get 100% accuracy which would constitute a case of overfitting and result in poor performance when applied to the test set. Thus, a validation set, which is independent from the training set, is used for parameter selection. In this process, ANN researchers usually use 'Cross Validation'. Moreover, using many nodes for hidden layers is another problem. It causes of longer training time and increased complexity. In addition, our two models both used MLP architectures as 1:25:1 and 1:5:5:1 which are simpler than the MLP architectures which are changing between 1:100:100:1 and 1:250:250:1. The following reasons explain why we have found better results in the more simplest ANN architecture;
1- to use unsupervised PCA for selecting meaningful data set from large data size before ANN classifier
2- to use better activation function (sigmoid is more effective then rectified linear function).
3- to select optimum momentum coefficient and learning rate

## CONCLUSIONS

Our results show that pattern recognition over UV-Vis spectra can provide substantial insights of light-GNS interactions. PCA of spectral database reveals the Fano resonance characteristics of the SPP. PCA allows us to observe the 3D vector fields of the spectral data which exhibits the homoclinic strange attractor fixed to resonant saddle points. LDA of spectra reveals the quantum confinement effects, and ion and electron oscillations. Such capabilities show that GNPs have high potentials in high energy density physics and fusion applications besides medical and industrial applications. After calculation of PCA coordinates, a nonlinear relation between the coordinates and the particle diameters is observed. It was due to the PCA coordinates that we could reduce the dimension of input data for the training of ANN, and thereby simplifying the inputs and outputs of ANN modeling.

In this study, we used four well-known and effective ML algorithms and found that the most error-free result came from ANN-2. ANN-2 is therefore, established as a powerful tool for estimating the diameters with high accuracy, since it resulted in the best recognition accuracy with a 100% by using 6 different diameters. In conclusion, a 4 layered ANN model could be suggested as a useful method for predicting the results of a proposed model in terms of saving time and cost, and that pattern recognition based multi hidden layered ANN is a powerful method for estimating the diameters of nanoparticles.

## ACKNOWLEDGEMENTS

Special thanks to Nanocomposix Inc. for providing TEM images and measurements of nanoparticles and MIE scattering calculator.